\documentclass[letterpaper]{article}
\usepackage{amsfonts}

\usepackage{graphicx}
\usepackage{flushend}
\usepackage{fullpage}
\usepackage{subfigure}
\usepackage{times}
\usepackage{ifpdf}
\usepackage{caption}
\usepackage{booktabs}
\DeclareCaptionType{copyrightbox}
\usepackage{url}
\usepackage{amsmath}

\renewcommand{\P}{\mathcal{P}}
\newcommand{\V}{\mathcal{V}}

\newcommand{\Enc}{\text{Enc}}

\newcommand{\sjs}{\ensuremath{F_2}}
\newcommand{\distinct}{\ensuremath{F_0}}

\usepackage{color}
\newcommand{\edit}[1]{{#1}}

\newcommand{\pmw}{{\sc PM}}
\newcommand{\matmult}{{\sc MatMult}}

\newcommand{\eat}[1]{}
\newcommand{\etal}{{\textit{et al.}}}

\title{Verifiable Computation with Massively Parallel Interactive Proofs}

\author{Justin Thaler\thanks{Harvard University, 
School of Engineering and Applied Sciences,
jthaler@seas.harvard.edu. Supported by the Department of Defense (DoD) through the National Defense Science \& Engineering Graduate Fellowship (NDSEG) Program, and in part by NSF grants CCF-0915922 and IIS-0964473.}
 \and 
 Mike Roberts \thanks{Harvard University, 
School of Engineering and Applied Sciences,
mroberts@seas.harvard.edu. This work was partially supported by the Intel Science and Technology Center for Visual Computing, NVIDIA, and the National Science Foundation under Grant No. PHY-0835713.}
\and
Michael Mitzenmacher\thanks{Harvard University, 
School of Engineering and Applied Sciences, michaelm@eecs.harvard.edu. This work was supported by NSF grants CCF-0915922 and IIS-0964473.}
\and Hanspeter Pfister \thanks{Harvard University, 
School of Engineering and Applied Sciences,
pfister@seas.harvard.edu. This work was partially supported by the Intel Science and Technology Center for Visual Computing, NVIDIA, and the National Science Foundation under Grant No. PHY-0835713.}}
\date{}

\renewcommand{\paragraph}[1]{\vspace{2mm}\noindent{\bf #1.}}

\begin{document}

\maketitle

\begin{abstract}
As the cloud computing paradigm has gained prominence, the need for \emph{verifiable computation} has grown increasingly urgent. The concept of verifiable computation enables a weak client to outsource difficult computations to a powerful, but untrusted, server. Protocols for verifiable computation aim to provide the client with a \emph{guarantee} that the server performed the requested computations correctly, without requiring the client to perform the requested computations herself. By design, these protocols impose a minimal computational burden on the client. However, existing protocols require the server to perform a very large amount of extra bookkeeping, on top of the requested computations, in order to enable a client to easily verify the results.  Verifiable computation has thus remained a theoretical curiosity, and protocols for it have not been implemented in real cloud computing systems.

In this paper, our goal is to leverage GPUs to reduce the server-side slowdown for verifiable computation. To this end, we identify abundant data parallelism in a state-of-the-art general-purpose protocol for verifiable computation, originally due to Goldwasser, Kalai, and Rothblum \cite{muggles}, and recently extended by Cormode, Mitzenmacher, and Thaler \cite{itcs}. We implement this protocol on the GPU, and we obtain 40-120$\times$ server-side speedups relative to a state-of-the-art sequential implementation. For benchmark problems, our implementation thereby reduces the slowdown of
the server to within factors of 100-500$\times$ relative to the original
computations requested by the client.  Furthermore, we reduce the already small runtime of the client by 100$\times$.
Similarly, we obtain 20-50$\times$ server-side and client-side speedups for related protocols targeted at specific streaming problems. We believe our results demonstrate the immediate practicality of using GPUs for verifiable computation, and more generally, that protocols for verifiable computation have become sufficiently mature to deploy in real cloud computing systems.

\end{abstract}

\section{Introduction}
A potential problem in outsourcing work to commercial cloud computing services is trust. If we store a large dataset with a server, and ask the server to perform a computation on that dataset -- for example, to compute the eigenvalues of a large graph, or to compute a linear program on a large matrix derived from a database -- how can we know the computation was performed correctly? Obviously we don't want to compute the result ourselves, and we might not even be able to store all the data locally. Despite these constraints, we would like the server to not only provide us with the answer, but to convince us the answer is correct. 

Protocols for \emph{verifiable computation} offer a possible solution to this problem. The ultimate goal of any such protocol is to enable the client to obtain results with a guarantee of correctness from the server much more efficiently than performing the computations herself. Another important goal of any such protocol is to enable the server to provide results with guarantees of correctness \emph{almost}  as efficiently as providing results without guarantees of correctness.

Interactive proofs are a powerful family of protocols for establishing guarantees of correctness between a client and server. Although they have been studied in the theory community for decades, there had been no significant efforts to implement or deploy such proof systems until very recently. A recent line of work (e.g., \cite{riva, icalp09, esa, itcs, vldb, muggles, ndss}) has made substantial progress in advancing the practicality of these techniques. In particular, prior work of Cormode, Mitzenmacher, and Thaler \cite{itcs} demonstrates that: (1) a powerful general-purpose methodology due to Goldwasser, Kalai and Rothblum \cite{muggles} approaches practicality; and (2) special-purpose protocols for a large class of streaming problems are already practical. 

In this paper, we clearly articulate this line of work to researchers outside the theory community. We also take things one step further, leveraging the parallelism offered by GPUs to obtain significant speedups relative to state-of-the-art implementations of \cite{itcs}. Our goal is to invest the parallelism of the GPU to obtain correctness guarantees with minimal slowdown, rather than to obtain raw speedups, as is the case with more traditional GPU applications. We believe the insights of our GPU implementation could also apply to a multi-core CPU implementation. However, GPUs are increasingly widespread, cost-effective, and power-efficient, and they offer the potential for speedups in excess of those possible with commodity multi-core CPUs \cite{owens, debunk}.

We obtain server-side speedups ranging from 40-120$\times$ for the general-purpose protocol due to Goldwasser \etal\ \cite{muggles}, and 20-50$\times$ speedups
for related protocols targeted at specific streaming problems. 
Our general-purpose implementation reduces the server-side cost of providing results with a guarantee of correctness to within factors of 100-500$\times$ relative to a sequential algorithm 
without guarantees of correctness. Similarly, our implementation
of the
special-purpose protocols reduces the server-side slowdown to within 10-100$\times$ relative to a sequential algorithm without guarantees of correctness.

We believe the additional costs of obtaining correctness guarantees demonstrated in this paper would already be considered modest in many correctness-critical applications. 
For example, at one end of the application spectrum is Assured Cloud Computing for military contexts: a military user may need integrity guarantees
when computing in the presence of cyber attacks, or may need such guarantees when coordinating critical computations
across a mixture of secure military networks and insecure networks owned by civilians or other nations \cite{airforce}.
At the other end of the spectrum, a hospital that outsources the processing of patients' electronic medical records to the cloud may require guarantees that the
server is not dropping or corrupting any of the records. Even if every computation is not explicitly checked, the mere ability to check the computation could 
mitigate trust issues and stimulate users to adopt cloud computing solutions.  

Our source code is available at \cite{code2}.
%We believe our results show that methods for verifiable computation have sufficiently evolved 
%to be worth deploying in real cloud-computing systems. 

%One of the goals of this paper is to clearly articulate this line of
%work to researchers outside of the theory community, and so we begin with a broad survey of the literature as it relates to this work.

\section{Background}

\subsection{What are interactive proofs?}
Interactive proofs (IPs) were introduced within the computer science theory community more than a quarter century ago,
in seminal papers by Babai \cite{ip1} and Goldwasser, Micali and Rackoff \cite{ip2}. In any IP,
there are two parties: a \emph{prover} $\P$, and a \emph{verifier} $\V$. $\P$ is typically considered to be computationally powerful, while $\V$ is
considered to be computationally weak. 

In an IP, $\P$ solves a problem using her (possibly vast) computational resources, and tells $\V$ the answer. $\P$ and $\V$ then
have a conversation, which is to say, they engage in a randomized protocol involving the exchange of one or more messages between the two parties.
The term \emph{interactive} proofs derives from the back-and-forth nature of this conversation.
During this conversation, $\P$'s goal is to convince $\V$ that her answer is correct. 

% (see Figure \ref{fig:ov} for a
%high-level depiction of such a protocol). 

IPs naturally model the problem of a client (whom we model as $\V$) outsourcing computation to an untrusted server (who we model as $\P$). 
That is, IPs provide a way for a client to hire a cloud computing service to store and process data, and to efficiently \emph{check} the integrity of the results returned by the server. This is useful whenever the server is not a trusted entity, either because the server is deliberately deceptive, or is simply buggy or inept. We therefore interchange the terms server and prover where appropriate. Similarly, we interchange the terms client and verifier where appropriate. 

Any IP must satisfy two properties. Roughly speaking, the first is that if $\P$ answers correctly and follows the prescribed protocol, then $\P$ will convince $\V$ to accept
the provided answer.
The second property is a security guarantee,
which says that if $\P$ is lying, then $\V$ must catch $\P$ in the lie and reject the provided answer with high probability. A trivial way to satisfy this property is to have $
\V$ compute the answer to the problem herself, and accept only if her answer matches $\P$'s. But this defeats the purpose of having a prover. The goal of an interactive 
proof system is to allow $\V$ to check $\P$'s answer using resources considerably smaller than those required to solve the problem from scratch. 

At first blush, this may appear difficult or even impossible to achieve. However, IPs have turned out to be surprisingly powerful. %, and a rich theory of their power has been developed. 
We direct the interested reader to \cite[Chapter 8]{arorabarak} for an excellent overview of this area.

\subsection{How do interactive proofs work?} 
\label{sec:10kft} At the highest level, many interactive proof methods (including the ones in this paper) work as follows.
Suppose the goal is to compute a function $f$ of the input $x$. 

First, the verifier makes a single streaming pass over the input $x$, during which she extracts a short \emph{secret} $s$. 
This secret 
is actually a single (randomly chosen) symbol of an error-corrected encoding $\Enc(x)$ of the input. To be clear, the secret does \emph{not} depend on the 
problem being solved; in fact, for many interactive proofs, it is not necessary that the problem be determined until \emph{after} the secret is extracted. 
% (called the \emph{low-degree extension} code in the 
%theory literature). 

\begin{figure}[t]
\includegraphics[width=1.0\linewidth]{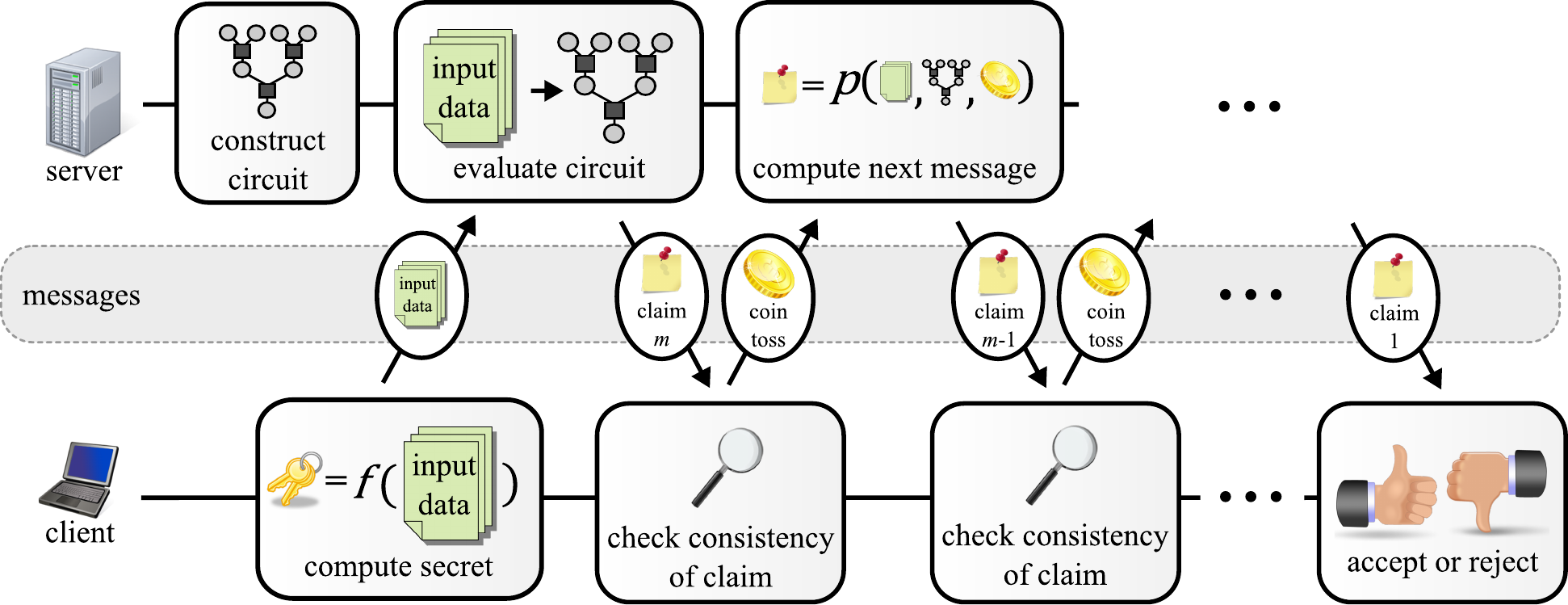}
\caption{High-level depiction of an execution of the GKR protocol.}
\label{fig:ov}
\end{figure}

Next, $\P$ and $\V$ engage in an extended conversation, during which
$\V$ sends $\P$
various challenges, and $\P$ responds to the challenges (see Figure \ref{fig:ov} for an illustration).
The challenges are all related to each other, 
and the verifier checks that the prover's responses to all challenges
are \emph{internally consistent}. 

The challenges are chosen so that the prover's response to the first challenge must include a (claimed) value for the function of interest.
Similarly, the prover's response to the last challenge must include 
a claim about what the value of the verifier's secret $s$ should be.
If all of $\P$'s responses are internally consistent, and the claimed value of $s$ matches the true value of $s$, then
the verifier is convinced that prover followed the prescribed protocol and accepts. Otherwise, the verifier
\emph{knows} that the prover deviated at some point, and rejects. From this point of view,
the purpose of all intermediate challenges is to guide the prover from a claim about $f(x)$ to a claim about the secret $s$,
while maintaining $\V$'s control over $\P$.

Intuitively, what gives the verifier surprising power to detect deviations is the error-correcting properties of $\Enc(x)$. 
Any good error-correcting code satisfies the property that if two strings $x$ and $x'$ differ in even one location,
then $\Enc(x)$ and $\Enc(x')$ differ in almost every location. In the same
way, interactive proofs ensure that if $\P$ flips even a single bit of a single message in the protocol, 
then $\P$ either has to make an inconsistent claim at some later point, or else has to lie almost everywhere in her final claim about the value
of the secret $s$.
Thus, if the prover deviates from the prescribed protocol \emph{even once} %in the protocol of \cite{muggles} 
%(say by flipping a single bit in a single message
%within the protocol), 
the verifier will detect this with high probability and reject.

\subsection{Previous work}
\label{sec:prevwork}

Unfortunately, despite their power, IPs have had very little influence on real systems where integrity guarantees on outsourced computation would be useful. There appears to have been a folklore 
belief that these methods are impractical \cite{ndss}.
As previously mentioned, a recent line of work (e.g., \cite{riva, icalp09, esa, itcs, vldb, muggles, ndss}) has made substantial progress in advancing the practicality of these techniques. 
In particular, Goldwasser \etal\ \cite{muggles} 
described a powerful general-purpose protocol (henceforth referred to as the GKR protocol) 
that achieves a polynomial-time prover and nearly linear-time verifier for a large class of computations. Very recently, Cormode, Mitzenmacher, and Thaler \cite{itcs} showed how to significantly speed up the prover in the GKR protocol \cite{muggles}. They also implemented this protocol, and demonstrated experimentally that their implementation approaches practicality. Even with their optimizations,
the bottleneck in the implementation of \cite{itcs} is the prover's runtime, with all other costs (such as verifier space and runtime) being extremely low.

A related line of work has looked at protocols for specific \emph{streaming} problems. Here, the goal is not just to save the verifier time (compared to doing 
the computation without a prover), but also to save the verifier \emph{space}. This is motivated by cloud computing settings where the client does not even have space to store a local copy of the input, and thus uses the cloud to both store and process the data.  The protocols developed in this line of work do not require the client to store the input, but rather allow the client to make a single streaming pass over the input (which can occur, for example, while the client is uploading data to the cloud). Throughout this paper, whenever we mention a \emph{streaming verifier}, we mean the verifier makes
a single pass over the input, and uses space significantly sublinear in the size of the data.%, and in particular does not have to store the full input.

The notion of a \emph{non-interactive}
streaming verifier was first put forth by Chakrabarti \etal\ \cite{icalp09}
and studied further by Cormode \etal\ \cite{esa}. These works allow the prover to
send only a single message to the verifier (e.g., as an attachment to an email, or posted on a website), with no communication
in the reverse direction. Moreover, these works present protocols achieving provably optimal tradeoffs between the size of the proof and 
the space used by the verifier for a variety of problems, ranging from matrix-vector multiplication to graph problems like bipartite perfect matching. %In this sense, the protocols of \cite{icalp09, esa} are \emph{non-interactive}.

Later, Cormode, Thaler, and Yi extended the streaming model of \cite{icalp09} to allow an \emph{interactive} prover and verifier, who actually have a conversation.
%the prover and verifier to actually have a conversation (i.e. they allowed \emph{interaction} between the parties.) 
They demonstrated that interaction allows for much more efficient
protocols in terms of client space, communication, and server running time than are possible in the one-message model of \cite{icalp09, esa}. 
It was also observed in this work that
the general-purpose GKR protocol works with just a streaming verifier. 
Finally, the aforementioned work of Cormode, Thaler, and Mitzenmacher \cite{itcs}
also showed how to use sophisticated Fast Fourier Transform (FFT) techniques to drastically speed up the prover's computation in 
the protocols of \cite{icalp09, esa}. 

Also relevant is work by Setty \etal\ \cite{ndss}, who implemented a
protocol for verifiable computation
due to Ishai \etal\ \cite{ishai}.
To set the stage for our results using parallelization,
in Section~\ref{sec:expts} we compare our approach with \cite{ndss} and \cite{itcs} in detail.  
As a summary, the implementation of the GKR protocol described
in both this paper and in \cite{itcs} has
several advantages over \cite{ndss}.
For example, the GKR implementation saves space and time for the
verifier even when outsourcing a single computation,
while \cite{ndss} saves time for the verifier only when batching together
several dozen computations at once and amortizing the verifier's cost
over the batch. Moreover, the GKR protocol is unconditionally secure against
computationally unbounded adversaries who deviate from the prescribed protocol, 
while the Ishai \etal\ protocol relies on cryptographic assumptions to obtain security guarantees. We present experimental results demonstrating
that that the prover in the sequential implementation of \cite{itcs} based on the
GKR protocol runs significantly faster than the prover in the implementation of
\cite{ndss} based on the Ishai \etal\ protocol \cite{ishai}. 

Based on this comparison, we use the sequential implementation of \cite{itcs} as our baseline.
We then present results that our new GPU-based implementation runs 40-120$\times$
faster than the sequential implementation in \cite{itcs}.

\section{Our interactive proof protocols}
\label{sec:overview}
In this section, we give an overview of the methods implemented in this paper. % ,and explain the insights necessary 
%to parallelize the computation of both the prover and the verifier. %We direct the interested reader to \cite{muggles, itcs, icalp09, esa}
%for full descriptions of these protocols. 
Due to their highly technical nature, we seek only to convey a high-level description of the protocols relevant to this paper, and
deliberately avoid rigorous definitions or theorems. We direct the interested reader to prior work for further details \cite{icalp09, esa, itcs, muggles}.

\begin{figure}[t]
\begin{center}
\includegraphics[width=0.35\linewidth]{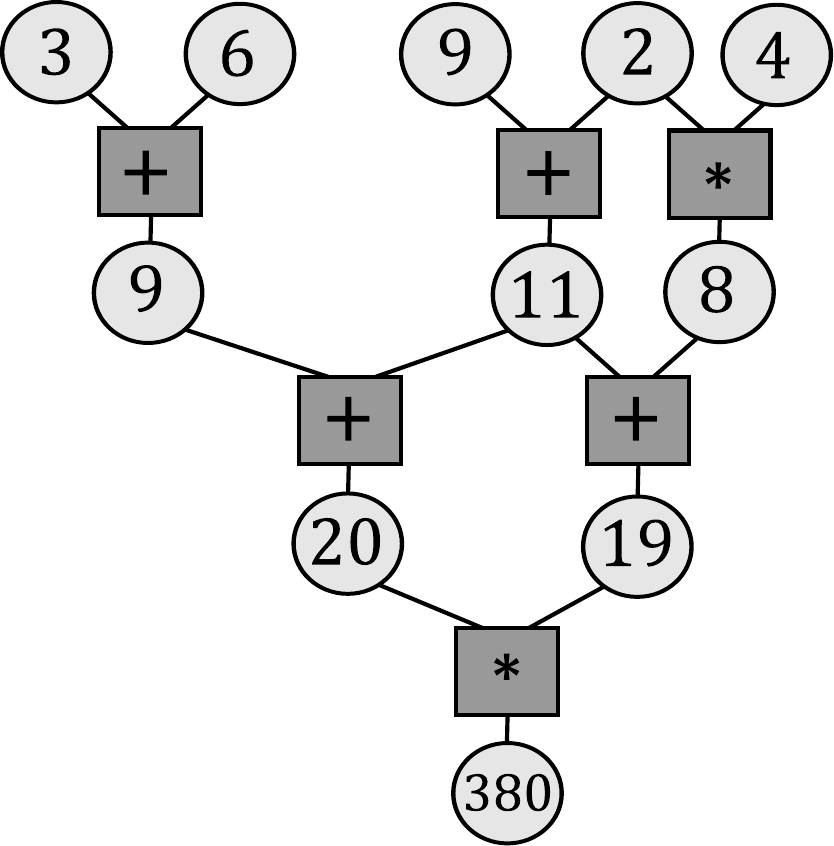}
\end{center}
\caption{A small arithmetic circuit.}
\label{fig:samplecirc}
\end{figure}

\subsection{GKR protocol}
\label{sec:mugglesdescription}

The prover and verifier first agree
on a layered arithmetic circuit of fan-in two over a finite field $\mathbb{F}$ computing the function of interest. An arithmetic circuit is just like a boolean circuit, 
except that the inputs are elements of $\mathbb{F}$ rather than boolean values, and the gates perform addition and multiplication
over the field $\mathbb{F}$, rather than computing $\text{AND}$, $\text{OR}$, and $\text{NOT}$ operations. See Figure \ref{fig:samplecirc} for an example circuit. In fact, any boolean circuit
can be transformed into an arithmetic circuit computing an equivalent function over a suitable finite field, although this approach may not yield the most succinct arithmetic circuit for the function.

Suppose the output layer of the circuit is layer $d$, and the input layer is layer $0$.
The protocol of \cite{muggles} proceeds in iterations, with one iteration for each layer of the circuit. The first iteration follows the 
general outline described in Section \ref{sec:10kft}, with $\V$ guiding $\P$ from a claim about the output of the circuit
to a claim about a secret $s$, via a sequence of challenges and responses. The challenges sent by $\V$ to $\P$ are simply
random coins, which are interpreted as random points in the finite field $\mathbb{F}$. The prescribed responses of $\P$ 
are polynomials, where each prescribed polynomial depends on the preceding challenge. Such a polynomial 
can be specified either by listing its coefficients,
or by listing its evaluations at several points. 

However, unlike in Section \ref{sec:10kft}, the secret $s$ is not a symbol in an error-corrected encoding of \emph{the input},
but rather a symbol in an error-corrected encoding of the \emph{gate values} at layer $d-1$. 
Unfortunately, $\V$ cannot 
compute this secret $s$ on her own. Doing so would require evaluating all previous layers of the circuit, and the whole point of outsourcing
is to avoid this. So $\V$ has $\P$
\emph{tell her} what $s$ should be. But now $\V$ has to make sure that $\P$ is not lying about $s$. 

This is what the second iteration accomplishes,
with $\V$ guiding $\P$ from a claim about $s$, to the claim about a new secret $s'$, which is a symbol in an encoding of the gate values at layer $d-2$.
This continues until we get to the input layer. At this point, the secret is actually a symbol in an error-corrected encoding of \emph{the input},
and $\V$ can compute this secret in advance from the input easily on her own. Figure \ref{fig:ov} illustrates the entirety of the GKR protocol at a very high level.

We take this opportunity to point out an important property of the protocol of \cite{muggles}, which was critical in allowing our GPU-based implementation
to scale to large inputs. Namely, \emph{any iteration of the protocol involves only two layers of the circuit at a time.} 
In the $i$th iteration, the verifier guides the prover from a claim about gate values at layer $d-i$ to a claim about gate values at layer $d-i-1$. 
Gates at higher or lower layers do not affect the prescribed responses within iteration $i$.

\subsection{Special-purpose protocols}

As mentioned in Section \ref{sec:prevwork}, efficient problem-specific \emph{non-interactive} verifiable protocols have been developed for a variety
of problems of central importance in streaming and database processing, ranging from linear programming to 
graph problems like shortest $s-t$ path. %An important property of many of these protocols is that they
%achieve \emph{optimal} tradeoffs between the length of the proof and the space used by the verifier. 
The central primitive in many of these protocols is itself a protocol originally due to Chakrabarti \etal\ \cite{icalp09}, 
for a problem known as the \emph{second frequency moment}, or $F_2$.
In this problem, the input is a sequence of $m$ items from a universe $\mathcal{U}$ of size $n$, and the goal is to compute
$F_2(x) = \sum_{i \in \mathcal{U}} f_i^2$, where $f_i$ is the number of times item $i$ appears in the sequence. As explained in \cite{itcs},
speeding up this primitive immediately speeds up protocols for all of the problems that use the $F_2$ protocol as a subroutine.

The aforementioned  $F_2$ protocol of Chakrabarti \etal\ \cite{icalp09} achieves provably \emph{optimal} tradeoffs between the 
length of the proof and the space used by the verifier. Specifically, for \emph{any} positive integer $h$, the protocol can achieve a proof length of just $h$ machine words, as long as the verifier
uses $v=O(n/h)$ words of space. For example, we may set both $h$ and $v$ to be roughly $\sqrt{n}$, which is substantially sublinear 
in the input size $n$.

Very roughly speaking, this protocol follows the same outline as in Section \ref{sec:10kft}, except that in order to remove the
\emph{interaction} from the protocol, the verifier needs to compute a more complicated secret. Specifically, the verifier's secret $s$
consists of $v$ symbols in an error-corrected encoding of the input, rather than a single
symbol. 
To compute the prescribed proof, the prover has to evaluate $2n$ symbols
in the error-corrected encoding of the input. The key insight of \cite{itcs} is that these $2n$ symbols need not be computed independently
(which would require substantially superlinear time), but
instead can be computed in $O(n \log n)$ time using FFT techniques. More specifically, the protocol of \cite{itcs} partitions the universe into
a $v\times h$ grid, and it performs a sophisticated FFT variant known as the \emph{Prime Factor Algorithm} \cite{pfa} on each row of the grid. 
The final step of $\P$'s computation is to compute the sum of the squared entries for each \emph{column} of the (transformed) grid; these values form the
actual content of $\P$'s prescribed message.

\section{Parallelizing our protocols}
\label{sec:issues}

In this section, we explain the insights necessary 
to parallelize the computation of both the prover and the verifier for the protocols we implemented.

\subsection{GKR protocol}
\subsubsection{Parallelizing $\P$'s computation}

% TOO PROOFY. START WITH THE 2ND PARA.

In every one of $\P$'s responses in the GKR protocol,
the prescribed message from $\P$ is defined via a large sum over roughly $S^3$ terms, 
where $S$ is the size of the circuit,
and so computing this sum naively would take $\Omega(S^3)$ time.
Roughly speaking, Cormode \etal\ in \cite{itcs} observe that each gate of the circuit contributes to only a single term of this sum,
and thus this sum can be computed via a single pass over the relevant gates. 
The contribution of each gate to the sum can be computed in constant
time, and each gate contributes to logarithmically many messages over the course of the protocol. 
Using these observations carefully reduces $\P$'s runtime from $\Omega(S^3)$,
to $O(S \log S)$, where again $S$ is the circuit size.

\begin{figure}[t]
\begin{center}
\includegraphics[width=.78\linewidth]{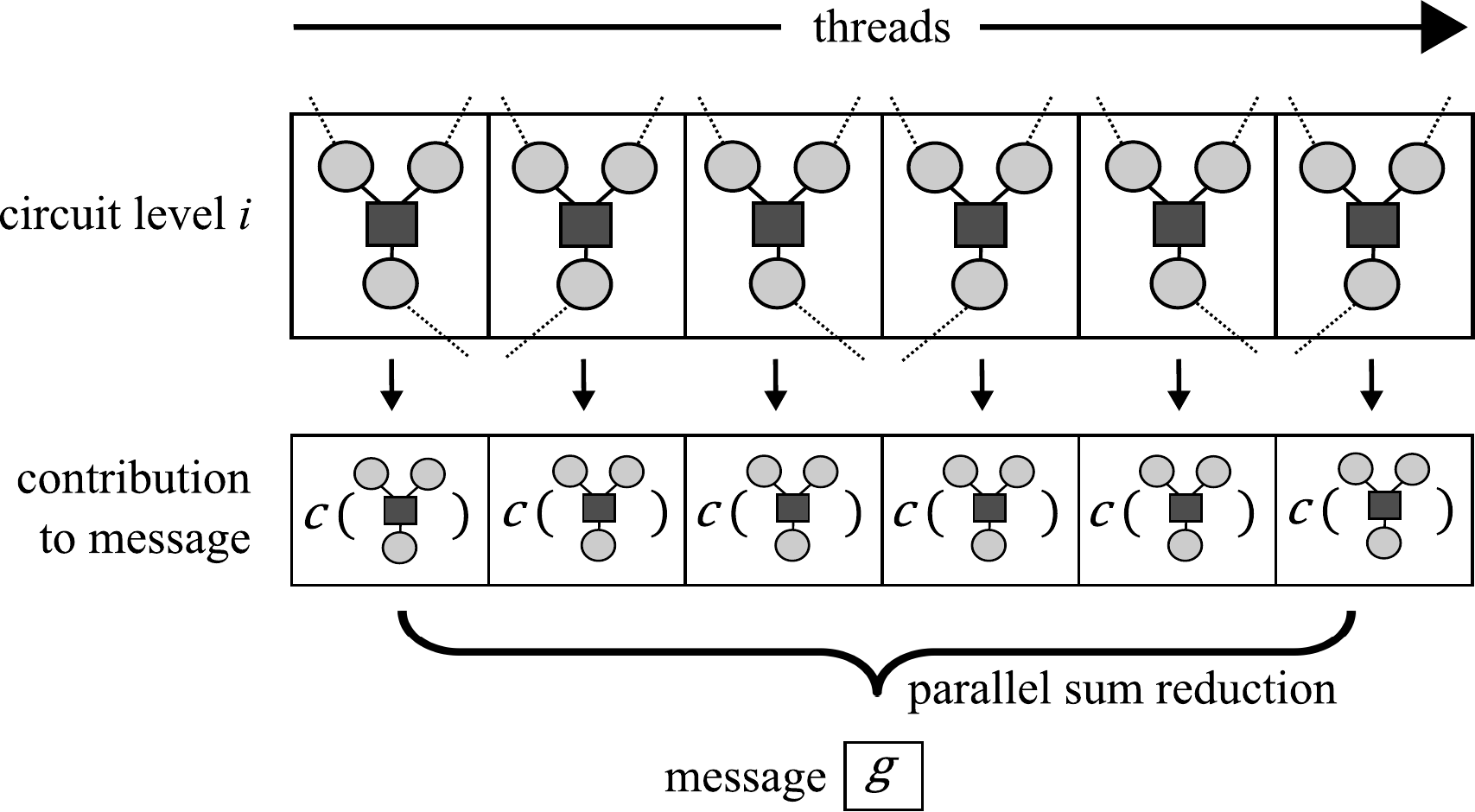}
\end{center}
\caption{Illustration of parallel computation of the server's message to the client in the GKR protocol.}
\label{fig:parmuggles}
\end{figure}

The same observation reveals that $\P$'s computation can be parallelized: each gate contributes \emph{independently} to the sum
in $\P$'s prescribed response. Therefore, $\P$ can compute the contribution of many gates in parallel, save the results in a temporary array,
and use a parallel reduction to sum the results. We stress that all arithmetic is done within the finite field $\mathbb{F}$, rather than over
the integers. Figure \ref{fig:parmuggles} illustrates this process.

\subsubsection{Parallelizing $\V$'s computation}
The bulk of $\V$'s computation (by far) consists of 
computing her secret, which consists of a single symbol $s$ in a particular error-corrected encoding of the input $x$. 
As observed in prior work \cite{vldb}, each symbol of the input contributes \emph{independently} to $s$. Thus, $\V$ can 
compute the contribution of many input symbols in parallel, and sum the results via a parallel reduction, just as in 
the parallel implementation of $\P$'s computation. This speedup is perhaps of secondary importance, as $\V$ runs extremely 
quickly even in the sequential implementation of \cite{itcs}. However, parallelizing $\V$'s computation is still an 
appealing goal, especially as GPUs are becoming  more common on personal computers and mobile devices.

\subsection{Special-purpose protocols}
\subsubsection{Parallelizing $\P$'s computation}
Recall that the prover in the special-purpose protocols
can compute the prescribed message by interpreting the input as a $v \times h$ grid, where $h$ is roughly the proof length and $v$ is the amount of space used by the verifier. The prover then performs a sophisticated FFT on each row of the grid independently.
This can be parallelized by transforming multiple
rows of the grid in parallel. Indeed, Cormode \etal~\cite{itcs} achieved roughly a 7$\times$ speedup 
for this problem by using all eight cores of a multicore processor. Here, we obtain a much larger 20-50$\times$ speedup using the GPU.
(Note that \cite{itcs} did not develop a parallel implementation of the GKR protocol, only of the special-purpose protocols).

\subsubsection{Parallelizing $\V$'s computation}
Recall that in the special-purpose protocols, the verifier's secret $s$
consists of $v$ symbols in an error-corrected encoding of the input, rather than a single
symbol. Just as in Section \ref{sec:mugglesdescription}, this computation can be parallelized by noting that each input symbol contributes
independently to each entry of the encoded input. This requires $\V$ to store a large buffer of input symbols to work on in parallel. In some streaming contexts, $\V$ may not have the memory to accomplish this. Still, there are many settings in which this is feasible. For example,
$\V$ may have several hundred megabytes of memory available, and seek to outsource processing of a stream that is many gigabytes or terabytes in 
length. Thus, parallel computation combined with buffering can help a streaming verifier keep up with a live stream of data: $\V$ splits her memory into two buffers, and at all times one buffer will be collecting arriving items. As long as $\V$ can process the full buffer (aided by parallelism) before her other buffer overflows, $\V$ will be able to keep up with the live data stream.  Notice this discussion applies to the client in the GKR protocol as well,
as the GKR protocol also enables a streaming verifier. %We demonstrate 

\section{Architectural considerations}
\label{sec:architecture}
\subsection{GKR protocol}

The primary issue with any GPU-based implementation of the prover in the GKR protocol is that the
computation is extremely
memory-intensive: for a circuit of size $S$ (which
corresponds to $S$ arithmetic operations in an
unverifiable algorithm), the prover in the GKR protocol has to store all $S$
gates explicitly, because she needs the
values of these gates to compute her prescribed messages. We investigate three alternative strategies for managing the memory overhead of the GKR protocol, which we refer to as the no-copying approach, the copy-once-per-layer approach, and the copy-every-message approach.

\subsubsection{The no-copying approach}

The simplest approach is to store
the entire circuit explicitly on the GPU. We call this the \emph{no-copying approach}. However, this means that the entire circuit must fit in device memory, a requirement which
is violated even for relatively small circuits, consisting of roughly tens of million of gates.  

\subsubsection{The copy-once-per-layer approach}
\label{sec:onceperlayer}
Another approach is to 
keep the circuit in host memory, and only copy information to the device when
it is needed. This is possible because, as mentioned in Section \ref{sec:mugglesdescription}, at any point in the protocol
the prover only operates on two layers of the circuit at a time, so 
only two layers of the circuit need to reside in device memory. We refer to this as the 
\emph{copy-once-per-layer approach}. This is the approach we used in 
the experiments in Section \ref{sec:expts}.

Care must be taken with this approach to prevent host-to-device copying from becoming a bottleneck. 
Fortunately, in the protocol for each layer there are several dozen messages to
be computed before the prover moves on to the next layer, and this ensures that 
the copying from host to device makes up a very small portion of the runtime.

This method is sufficient to scale to very large circuits for
all of the problems considered in the experimental section of \cite{itcs},
 since no single layer of the circuits is significantly larger than the problem input itself.
 However, this method remains problematic for circuits that have (one or several) layers
 which are particularly wide, as an explicit representation of all the gates within a \emph{single} wide layer may
 still be too large to fit in device memory.

\subsubsection{The copy-every-message approach} 
In the event that there are individual layers which are too large to reside in device memory, a third approach 
is to copy \emph{part} of a layer at a
time from the host to the device, and compute the contribution of each gate in the part to the prover's
message before swapping the part back to host memory and bringing in
the next part. We call this the copy-every-message approach. 
This approach is viable, but it raises a significant issue, alluded to in its name. Namely,
this approach requires host-to-device copying for \emph{every}
message, rather than just once per layer of the circuit. That is, in any iteration $i$ of the protocol, $\P$ cannot compute her $j$th message until after the $(j-1)$th challenge from $\V$
is received. Thus, for \emph{each message} $j$, the entirety of the $i$th layer must be loaded piece-by-piece into device memory, swapping each piece
back to host memory after the piece has been processed. 
In contrast, the copy-once-per-layer approach allows $\P$ to copy an entire layer $i$ to the device and leave the entire layer in device memory
for the entirety of iteration $i$ (which will consist of several dozen messages).  
Thus, the slowdown inherent in the copy-every-message approach is not just that
$\P$ has to break each layer into parts, but that $\P$ has to do host-to-device and device-to-host copying for each message,
instead of copying an entire layer and computing several messages from that layer.
% In contrast,
% if the whole layer fits in memory, $\P$ can do a single host-to-device copy for the entire layer, and compute several dozen messages before
% another host-to-device copy is necessary.

%copying must be done dozens of times
%more frequently than in the copy-once-per-layer approach.

We leave implementing the copy-once-per-message approach in full for future work,
but preliminary experiments %(done by having our copy-once-per-layer implementation perform unnecessary copying for each message) 
suggest that this approach is viable in practice, resulting in less than a 3$\times$ slowdown
compared to the copy-once-per-layer approach. Notice that even after paying this slowdown, our GPU-based implementation would still achieve 
a 10-40$\times$ speedup compared to the sequential implementation of \cite{itcs}.

\edit{\subsubsection{Memory access}
Recall that for each message in the $i$th iteration of the GKR protocol, we assign a thread to each gate $g$ at the $i$th layer of the circuit,
as each gate contributes independently to the prescribed message of the prover. The contribution of gate $g$
depends only on the index of $g$, the indices of the two gates feeding into $g$, and the \emph{values} of the two gates feeding into $g$.

Given this data, the contribution of gate $g$ to the prescribed message can be computed using roughly 10-20 additions and multiplications within the finite field $\mathbb{F}$
(the precise number of arithmetic operations required varies over the course of the iteration).
As described in Section \ref{sec:expts}, we choose to work over a field which allows for extremely efficient arithmetic; for example, multiplying two field elements requires three
machine multiplications of 64-bit data types, and a handful of additions and bit shifts.

In all of the circuits we consider, the \emph{indices} of $g$'s in-neighbors can be determined with very little arithmetic and no global memory accesses. For example, 
if the wiring pattern of the circuit forms a binary tree, then the first in-neighbor of $g$ has index $2\cdot\text{index}(g)$, and the second in-neighbor of $g$ has index $2 \cdot \text{index}(g)+1$.
For each message, the thread assigned to $g$ can compute this information from scratch without incurring any memory accesses.

In contrast, obtaining the \emph{values} of g's in-neighbors requires fetching 8 bytes per in-neighbor from global memory. Memory accesses are necessary because it is infeasible to compute the value of each gate's in-neighbors from scratch each message,
and so we store these values explicitly. As these global memory accesses can be a bottleneck in the protocol, we strive to arrange the data in memory to ensure that adjacent threads access adjacent memory locations.
To this end, for each layer $i$ we maintain two separate arrays, with the $j$'th entry of the first (respectively, second) array storing the first (respectively, second) in-neighbor of the $j$'th gate at layer $i$. 
During iteration $i$, the thread assigned to the $j$th gate accesses location $j$ of the first and second array to retrieve the value of its first and second in-neighbors respectively. This ensures that adjacent threads access adjacent memory locations. 

For all layers, the corresponding arrays are populated with in-neighbor values when we evaluate the circuit at the start of the protocol (we store each layer $i$'s arrays on the host until the $i$'th iteration of the protocol,
at which point we transfer the array from host memory to device memory as described in Section \ref{sec:onceperlayer}). Notice this methodology sometimes requires data duplication: if many gates at layer $i$ share the same in-neighbor $g_1$, then $g_1$'s value will appear many times in layer $i$'s arrays. We feel that slightly increased space usage is a reasonable price to pay to ensure memory coalescing.}

\subsection{Special-purpose protocols} 

\subsubsection{Memory access}
\label{sec:accesspatterns}
%One challenge associated with implementing our special-purpose protocols on the GPU, is that irregular data access patterns come at a significant cost.

Recall that the prover in our special-purpose protocols 
views the input as a $v \times h$ grid, and performs a sophisticated FFT on each row of the grid independently. Although the independence of calculations in each row offers abundant opportunities for \emph{task-parallelism}, extracting the \emph{data-parallelism} required for high performance on GPUs requires care due to the irregular memory access pattern of the specific FFT algorithm used.

We observe that although each FFT has a highly irregular memory access pattern, this memory access pattern is \emph{data-independent}. Thus, we can convert abundant task-parallelism into abundant data-parallelism by transposing the data grid into column-major rather than row-major order. This simple transformation ensures perfect memory coalescing despite the irregular memory access pattern of each FFT, and improves the performance of our special-purpose prover by more than 10$\times$.

\eat{\subsubsection{Scaling to large inputs} Compared to the GKR protocol, the special-purpose protocols 
are not memory-intensive -- in total, the prover requires space roughly four times the size of the input, compared to superlinear
space for the GKR protocol for any problem requiring superlinear time to solve. %Nonetheless, the limiting factor in the size of the inputs we can handle 
% is device memory, and we explored some methods to scale to larger inputs.

The special-purpose protocols have the appealing property that
the prover only needs to operate on a very small fraction of the input at a time. Specifically, after breaking the input into
a $\sqrt{n} \times \sqrt{n}$ grid, the prover's computation treats each row of the grid independently. If we store the grid in row-major order,
we can keep the GPU occupied by copying just a few thousand rows to the device at any time, keeping the remainder of the input in host memory. 
For this strategy to be compatible with the optimized memory access pattern described in in Section \ref{sec:accesspatterns}, we transpose our GPU working set into a column-major layout prior to performing each DFT.}

\section{Evaluation}
\label{sec:expts}

\subsection{Implementation details}
Except where noted, we performed our experiments on an Intel Xeon 3 GHz workstation with 16 GB of host memory. Our workstation also has an NVIDIA GeForce GTX 480 GPU with 1.5 GB of device memory. We implemented all our GPU code in CUDA and Thrust~\cite{thrust} with all compiler optimizations turned on. 

Similar to the sequential implementations of \cite{itcs}, both our implementation of the GKR protocol and the special-purpose $F_2$ protocol due to \cite{icalp09, itcs} work over the finite field $\mathbb{F}_p$ with $p=2^{61}-1$. We chose this field for a number of reasons. Firstly, the integers embed naturally within it. Secondly, the field is large enough that the probability the verifier fails
to detect a cheating prover is tiny (roughly proportional to reciprocal of the field size). Thirdly, arithmetic within the field can be performed efficiently with simple shifts and bit-wise operations \cite{thorup}.
\edit{We remark that we used no floating point operations were necessary in any of our implementations, because all arithmetic is done over finite fields. 

Finally, we stress that in all reported costs below, we do count the time taken to copy data between the host and the device, and all reported speedups relative to sequential processing 
take this cost into account. We do not count the time to allocate memory for scratch space because this can be done in advance.}

\subsection{Experimental methodology for the GKR protocol} We ran our GPU-based implementation of the GKR protocol on four separate circuits, 
which together capture several different aspects of computation, from data aggregation to search, to linear algebra. %All four are natural and well-motivated benchmarks.
The first three
circuits were described and evaluated in \cite{itcs} using the sequential
implementation of the GKR protocol. The fourth problem was described and evaluated in \cite{ndss} based on the Ishai \etal\ protocol \cite{ishai}. Below, $[n]$ denotes the integers $\{0, 1, \dots, n-1\}$.

\begin{itemize}
\item $F_2$: Given a stream of $m$ elements from $[n]$, compute
$\sum_{i\in [n]}  a_i^2$ where $a_i$ is the number of occurrences of $i$ in
the stream.  
\item \distinct:
Given a stream of $m$ elements from $[n]$, compute
the number of {\em distinct} elements (i.e., 
the number of $i$ with $a_i \neq 0$, where again $a_i$ is the number of
occurrences of $i$ in the stream). 
\item
\pmw:
%{\sc Pattern-Matching with Wildcards:} 
Given a stream representing text $T = (t_0, \dots , t_{n-1}) \in
[n]^n$ and pattern $P= (p_0, \dots , p_{q-1})\in[n]^q$, 
the pattern $P$ is said to
occur at location $i$ in $t$ if, for every position $j$ in $P$, $p_j = t_{i+j}$.
 The pattern-matching problem
is to determine the number of locations at which $P$ occurs in $T$. 
\item \matmult: Given three matrices $A$, $B$, $C \in [n]^{m^2}$, determine whether $AB=C$.
(In practice, we do not expect $C$ to truly be part of the input data stream. Rather, prior work \cite{vldb, itcs} has shown that 
the GKR protocol works even if $A$ and $B$ are specified from a stream, while $C$ is given later by $\P$.)
\end{itemize}

The first two problems, $F_2$ and \distinct, are classical data aggregation queries which have been studied for 
more than a decade in the data streaming community. \distinct\ is also a highly useful subroutine in more complicated
computations, as it effectively allows for equality testing of vectors or matrices (by subtracting two vectors and seeing
if the result is equal to the zero vector). We make use of this subroutine when designing our matrix-multiplication circuit below.

The third problem, \pmw, is a classic search problem, and is motivated, for example, by clients wishing to store (and search) their email on the cloud. Cormode \etal\ \cite{itcs} considered the {\sc Pattern Matching with Wildcards} problem, where
the pattern and text can contain wildcard symbols that match with any character, but for simplicity we did not implement this additional functionality.

We chose the fourth problem, matrix multiplication, for several reasons. First was its practical
importance. Second was a desire to experiment on problems requiring super-linear time to solve (in contrast to $F_2$ and $F_0$): running on a super-linear problem allowed us to demonstrate that our implementation as well as that of \cite{itcs}
saves the verifier \emph{time} in addition to space, and it also forced us to grapple with the memory-intensive nature of the GKR protocol (see Section \ref{sec:issues}).
Third was its status as a benchmark enabling us to compare the implementations of \cite{itcs} and \cite{ndss}. Although there are also efficient special-purpose protocols
to verify matrix multiplication (see Freivald's algorithm \cite[Section 7.1]{randombook}, as well as Chakrabarti \etal\ \cite[Theorem 5.2]{icalp09}), it is still interesting to see how a general-purpose
implementation performs on this problem. Finally, matrix multiplication is an attractive primitive to have at one's disposal when verifying more complicated
computations using the GKR protocol.

\begin{figure}[t]
\begin{center}
\includegraphics[width=0.45\linewidth]{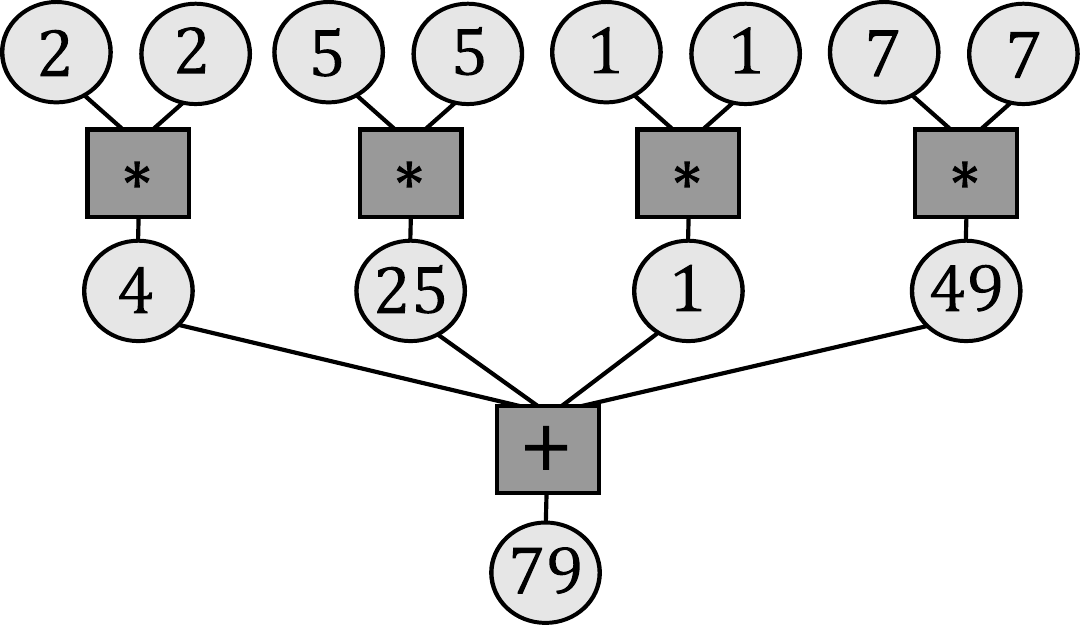}
\end{center}
\caption{The circuit for $F_2$.}
\label{fig:F2circ}
\end{figure}

\subsubsection{Description of circuits}
We briefly review the circuits for our benchmark problems.

The circuit for $F_2$ is by far the simplest (see Figure \ref{fig:F2circ} for an illustration). This circuit simply computes the square of each input wire using a layer of multiplication gates, and then sums the results
using a single sum-gate of very large fan-in. We remark that the GKR protocol typically assumes that all gates have fan-in two, but \cite{itcs} explains
how the protocol can be modified to handle a single sum-gate of very large fan-in at the output. 

The circuit for $F_0$ exploits Fermat's Little Theorem, which says that for prime $p$, $a^{p-1} \equiv 1 \mod p $ if and only if $a \neq 0$. Thus, this circuit computes
the $p-1$'th power of each input wire (taking all non-zero inputs to 1, and leaving all 0-inputs at 0), and sums the results via a single sum-gate of high fan-in.

The circuit for \pmw\ is similar to that for $F_0$: essentially, for each possible location of the pattern, it computes a value that is 0 if the pattern is at the location, and non-zero otherwise. It then computes the $(p-1)$th power of each such value and sums the results (i.e., it uses the $F_0$ circuit as a subroutine) to determine the number of locations where the pattern does (not) appear in the input.

Our circuit for \matmult\ uses similar ideas. We could run a separate instance of the GKR protocol to verify each of the $n^2$ entries in the output matrix $AB$ and compare them to $C$, but this
would be very expensive for both the client and the server. Instead, we specify a suitable circuit with a \emph{single} output gate, allowing us to run a single instance of the protocol to verify the output. Our circuit computes the $n^2$ entries in $AB$ via naive matrix multiplication, and subtracts the corresponding entry of $C$ from each. It then computes the number of non-zero values using the $F_0$ circuit as a subroutine. The final output of the circuit is zero if and only if $C = AB$.

\subsubsection{Scaling to large inputs}
As described in Section \ref{sec:architecture}, the memory-intensive nature of the GKR protocol made it challenging to scale to large inputs, especially given the
limited amount of device memory. Indeed, with the no-copying approach (where we simply keep the entire circuit in device memory), we were only able to scale to inputs of size roughly $150,000$ for the $F_0$ problem, and to $32 \times 32$ matrices for the \matmult\ problem on a machine with 1 GB of device memory. Using the copy-once-per-layer approach, we were able to scale to inputs with over 2 million entries for the $F_0$ problem, and $128 \times 128$ matrices for the \matmult\ problem. By running on a NVIDIA Tesla C2070 GPU with 6 GBs of device memory, we were able to push to $256 \times 256$ matrices
for the \matmult\ problem; the data from this experiment is reported in Table \ref{tabmuggles}.

\subsubsection{Evaluation of previous implementations}
\label{sec:comparison}
To our knowledge, the only existing implementation for verifiable computation that can be directly compared to that of Cormode \etal\ \cite{itcs} is  that
of Setty \etal\ \cite{ndss}. We therefore performed a brief comparison of the sequential implementation of \cite{itcs} with that of \cite{ndss}. This provides important context 
in which to evaluate our results: our 40-120$\times$ speedups compared to the sequential implementation of \cite{itcs} would be 
less interesting if the sequential implementation of \cite{itcs} was slower than alternative methods. Prior to this paper, these implementations had never been run on the same problems, so we picked a benchmark problem (matrix multiplication) evaluated in \cite{ndss} and compared to the results
reported there.

We stress that our goal is not to provide a rigorous quantitative comparison of the two implementations. Indeed, we only compare the implementation of \cite{itcs} to the numbers reported in \cite{ndss}; we never ran
the implementations on the same system, leaving this more rigorous comparison for future work. Moreover, both implementations may be amenable to further optimization. Despite these caveats, the comparison between the two implementations seems clear. The results are summarized in Table \ref{tab:comp}. 
\begin{table*}[t]
\centering
\begin{tabular}{ l l l l l l l }
%\begin{centering}
\toprule
Implementation & Matrix Size & $\P$ Time & $\V$ Time & Total Communication \\  
\midrule
\cite{itcs}                  & $512 \times 512$ & 3.11 hours        & 0.12 seconds         & 138.1 KB          \\
\cite{ndss}, \text{Pepper}   & $400 \times 400$ & 8.1 years$^\ast$  & 14 hours$^\ast$       & Not Reported      \\
\cite{ndss}, \text{Habanero} & $400 \times 400$ & 17 days$^\dagger$ & 2.1 minutes$^\dagger$ & 17.1 GB$^\dagger$ \\
\bottomrule
\end{tabular}
\centering
\caption{Comparison of the costs for the sequential implementations of \cite{itcs} and \cite{ndss}. Entries marked with $^\ast$ indicate that the costs given are \emph{total} costs over 45,000 queries.
Entries marked with $^\dagger$ indicate that the costs are total costs over 111 queries.}
\label{tab:comp}
\end{table*}

In Table \ref{tab:comp}, \text{Pepper} refers to an implementation in \cite{ndss} which is actually proven secure against polynomial-time adversaries under cryptographic assumptions, while \text{Habenero} is an implementation in \cite{ndss} which runs faster by allowing for a very high soundness probability of $\frac{7}{9}$ that a deviating prover can fool the verifier, and utilizing what the authors themselves refer to as heuristics (not proven secure in \cite{ndss}, though the authors indicate this may be due to space constraints). In contrast, the soundness probability in the implementation of \cite{itcs} is roughly $\frac{1}{2^{50}}$ (roughly proportional to the reciprocal of the field size $p=2^{61}-1$), and the protocol is unconditionally secure even against computationally unbounded adversaries. 

The implementation of \cite{ndss} has very high set-up costs for both $\P$ and $\V$, and therefore the costs of a \emph{single} query are very high. But this set-up cost can be amortized over many queries, and the most detailed experimental results provided in \cite{ndss} give the costs for batches of hundreds or thousands of queries. The costs reported in the second and third rows of Table \ref{tab:comp} are therefore the \emph{total} costs of the implementation when run on a large number of queries. 

When we run the implementation of \cite{itcs} on a single $512 \times 512$ matrix, the server takes 3.11 hours, the client takes 0.12 seconds, and the total length of all messages transmitted between the two parties is 138.1 KB. In contrast, the server in the heuristic implementation of \cite{ndss}, \text{Habanero}, requires 17 days amortized over 111 queries when run on considerably smaller matrices ($400 \times 400)$. This translates to roughly $3.7$ hours per query, but the cost of a single query without batching is likely about two orders of magnitude higher. The client in \text{Habanero} requires 2.1 minutes to process the same 111 queries, or a little over 1 second per query, while the total communication is 17.1 GBs, or about 157 MBs per query. Again, the per query costs will be roughly two orders of magnitude higher without the batching. 

We conclude that, even under large batching the per-query time for the server of the sequential implementation of \cite{itcs} is competitive with the heuristic implementation of \cite{ndss}, while the per-query time for the verifier is about two orders of magnitude smaller, and the per-query communication cost is between two and three orders of magnitude smaller. Without the batching, the per-query time of \cite{itcs} is roughly 100$\times$ smaller for the server and 1,000$\times$ smaller for the client, and the communication cost is about 100,000$\times$ smaller.
 
 Likewise, the implementation of \cite{itcs} is over 5 orders of magnitude faster for the client than the non-heuristic implementation \text{Pepper}, and four orders of magnitude faster for the server.

\begin{figure*}[t]
\begin{center}
\includegraphics[width=1.0\linewidth]{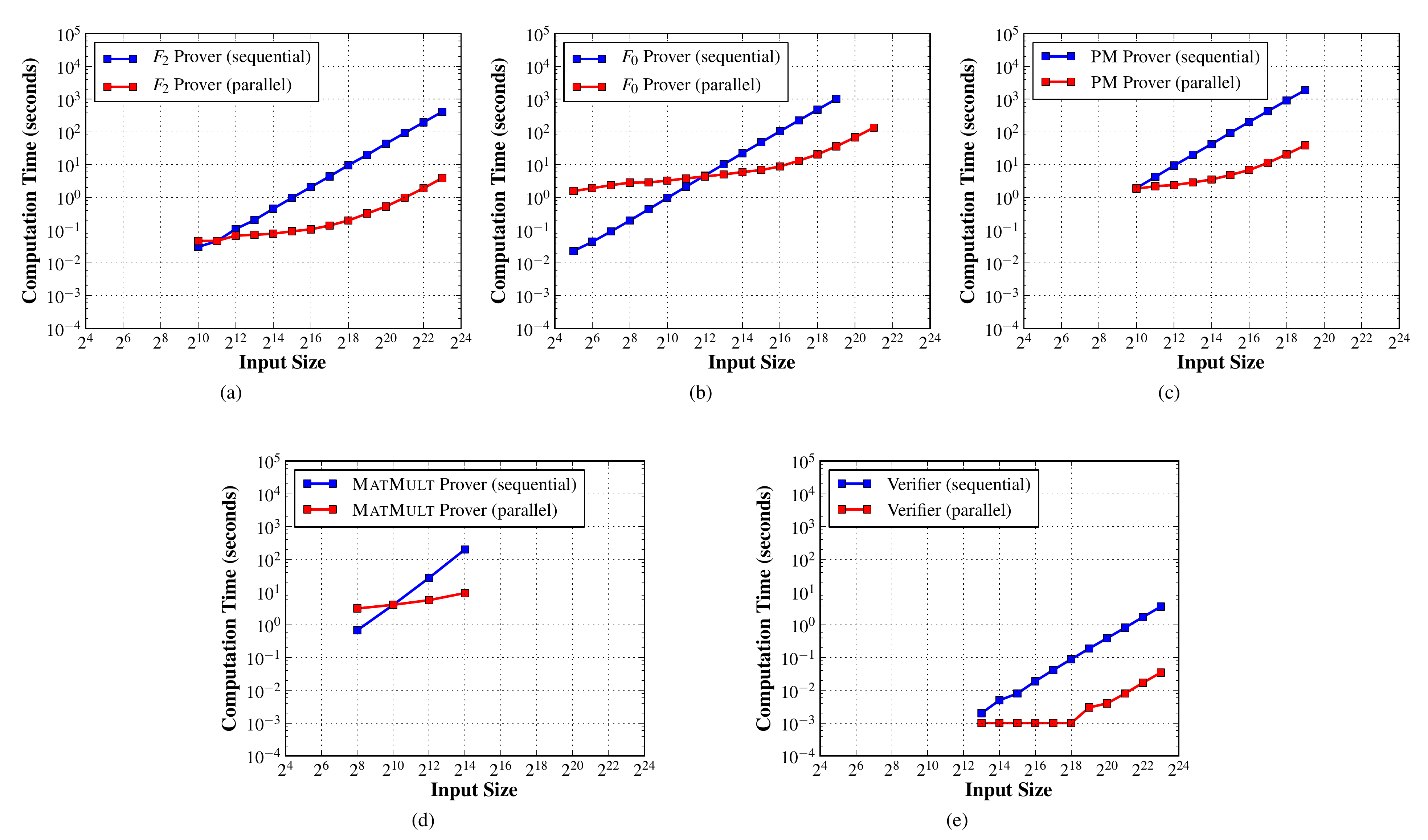}
\end{center}
\caption{Comparison of prover and verifier runtimes between the sequential implementation of the GKR protocol due to \cite{itcs} and our GPU-based implementation. 
Note that all plots are on a log-log scale. Plots (a), (b), (c), and (d)  depict the prover runtimes for $F_0$, $F_2$, \pmw, \matmult $ $ respectively. Plot (e) depicts the verifier runtimes for the GKR protocol. We include only one plot for the verifier, since its dominant cost in the GKR protocol
is problem-independent.}
\label{fig:mugglesresults}
\end{figure*}

\subsubsection{Evaluation of our GPU-based implementation} 

\begin{table*}[t]
\small
\centering
\begin{tabular}{ @{}l l l l l l l l l } 
%\begin{centering}
\toprule
Problem & Input Size   & Circuit Size & GPU $\P$ & Sequential     & Circuit     & GPU $\V$ & Sequential    & Unverified \\
        & (number of    & (number of   & Time (s) & $\P$ Time (s)  & Evaluation  & Time (s) & $\V$ Time (s) & Algorithm  \\
        & entries)             & gates)       &          &                & Time (s)    &          &               & Time (s)   \\
\midrule
\sjs      &  8.4 million     & 25.2 million     & 3.7     & 424.6   &  0.1 & 0.035 &       3.600 & 0.028    \\
\distinct &  2.1 million     & 255.8 million     & 128.5     & 8,268.0 &  4.2 &0.009 & 0.826   & 0.005    \\
\pmw      & 524,288           & 76.0 million & 38.9      & 1,893.1 & 1.2 &  0.004 & 0.124 & 0.006 \\
\matmult  & 65,536              & 42.3 million     & 39.6              & 1,658.0 & 0.9 & 0.003 & 0.045 & 0.080 \\
\midrule
\end{tabular}
\caption{ Prover runtimes in the GKR protocol for all four problems considered.}
%% $L+$ denotes an addition gate with large fan-in.}
\label{tabmuggles}
%\vspace*{-3mm}
\end{table*}

Figure \ref{fig:mugglesresults} demonstrates the performance of our GPU-based implementation of the GKR protocol. Table \ref{tabmuggles} also gives a succinct summary of our results, showing the costs for the largest instance of each problem we ran on. We consider the main takeaways of our experiments to be the following.

\medskip \noindent \textbf{Server-side speedup obtained by GPU computing.}
Compared to the sequential implementation of \cite{itcs}, our GPU-based server implementation ran close to 115$\times$ faster for the $F_2$ circuit, about
60$\times$ faster for the $F_0$ circuit, 45$\times$ faster for \pmw, and about 40$\times$ faster for \matmult $ $ (see Figure \ref{fig:mugglesresults}). 

Notice that for the first three problems, we need to look at large inputs to see the asymptotic behavior of the curve corresponding to the parallel prover's runtime. Due to the log-log scale in Figure \ref{fig:mugglesresults}, the curves for both the sequential and parallel implementations are asymptotically linear, and the 45-120$\times$ speedup obtained by our GPU-based implementation
is manifested as an additive gap between the two curves.
The explanation for this is simple: there is considerable overhead relative to the total computation time in parallelizing the computation
at small inputs, but this overhead is more effectively amortized as the input size grows. %At sufficiently large inputs, the overhead is negligible, and 
%both the parallel and sequential implementations' runtime grows roughly linearly with the input size (though the leading constant is 40x-120x smaller for the parallel implementation). 

In contrast, notice that for \matmult\ the slope of the curve for the parallel prover remains significantly smaller than that of the sequential prover throughout the entire plot. This is because
our GPU-based implementation ran out of device memory
well before the overhead in parallelizing the prover's computation became negligible. 
We therefore believe the speedup for \matmult\ would be somewhat higher than the 40$\times$ speedup observed if we were able to run on larger inputs.

\medskip 
\noindent \textbf{Could a parallel verifiable program be faster than a sequential unverifiable one?} 
The very first step
of the prover's computation in the GKR protocol is to evaluate the circuit. In theory this can be done efficiently in parallel, by proceeding sequentially layer by layer and 
evaluating all gates at a given layer in parallel. However, in practice we observed that the time it takes to copy the circuit to the device exceeds the time it takes to evaluate
the circuit sequentially.  This observation suggests that on the current generation of GPUs, no GPU-based implementation of the prover
could run \emph{faster} than a sequential \emph{unverifiable} algorithm. This is because sequentially evaluating
 the circuit takes at least as long as the unverifiable sequential algorithm, and copying the data to the GPU takes longer than sequentially evaluating the circuit. This observation
 applies not just to the GKR protocol, but to any protocol that uses a circuit representation of the computation (which is a standard technique in the theory literature \cite{ishai, hotos}).
Nonetheless, we can certainly hope to obtain a GPU-based implementation that is \emph{competitive} with sequential
unverifiable algorithms.

\medskip \noindent \textbf{Server-side slowdown relative to unverifiable sequential algorithms.}
For $F_2$, the total slowdown for the prover was roughly 130$\times$ (3.7 seconds compared to 0.028 seconds for the unverifiable algorithm,
which simply iterates over all entries of the frequency vector and computes the sum of the squares of each entry). We stress
that it is likely that we overestimate the slowdown resulting from our protocol, because we did not count the time it takes for 
the unverifiable implementation to compute the number of occurrences of each item $i$, that is, to \emph{aggregate} the stream into its frequency vector representation $(a_1, \dots, a_n)$. Instead, we simply generated the vector of frequencies at random (we did not count the generation time),
and calculated the time to compute the sum of their squares.
In practice, this aggregation step
may take much longer than the time required to compute the sum of the squared frequencies once the stream is in aggregated form.
%Certainly for streams whose length $m$ is significantly longer than the universe size $n$, the slowdown from our protocol could fall by an order of magnitude or more.

For \distinct, our GPU-based server implementation ran roughly 25,000$\times$ slower than the obvious unverifiable algorithm which simply counts the number of non-zero items in a vector. The larger slowdown compared to the $F_2$ problem is unsurprising. Since \distinct\ is a less arithmetic problem
than $F_2$, its circuit representation is much larger. Once again, it is likely that we overestimate the slowdowns for this problem, as we did not count the time for an unverifiable algorithm to
aggregate the stream into its frequency-vector representation. Despite the substantial slow-down incurred for $F_0$ compared to a naive unverifiable algorithm, it remains valuable as a 
primitive for use in heavier-duty computations like \pmw\ and \matmult.

For \pmw, the bulk of the circuit consists of a \distinct\ sub-routine, and so the runtime of our GPU-based implementation was similar to those for \distinct. However,
the sequential unverifiable algorithm for \pmw\ takes longer than that for \distinct.
 Thus, our GPU-based server implementation ran roughly 6,500$\times$ slower than the naive unverifiable algorithm, which exhaustively searches all possible locations for occurrences of the pattern.

For \matmult, our GPU-based server implementation ran roughly 500$\times$ slower than naive matrix-multiplication for $256 \times 256$ matrices. 
Moreover, this number is likely inflated due to cache effects from which the naive unverifiable algorithm benefited. That is, the naive unverifiable algorithm takes only $0.09$ seconds for $256 \times 256$ matrices, but takes $7.1$ seconds for $512 \times 512$ matrices, likely because the algorithm experiences very few cache misses on the smaller matrix. We therefore expect the slowdown of our implementation to fall to under 100$\times$ if we were to scale to larger matrices. 
Furthermore, the GKR protocol is capable of verifying matrix-multiplication \emph{over the finite field $\mathbb{F}_p$} rather than over the integers at no additional cost. Naive matrix-multiplication over this field is between 2-3$\times$ slower than matrix multiplication over the integers (even using the fast
arithmetic operations available for this field). Thus, if our goal was to work over this finite field rather than the integers, our slowdown would fall by another 2-3$\times$. It is therefore possible that our server-side slowdown may be less than 50$\times$ at larger inputs compared to naive matrix multiplication over $\mathbb{F}_p$.

\medskip \noindent \textbf{Client-side speedup obtained by GPU computing.}
The bulk of $\V$'s computation consists of evaluating a single symbol in an error-corrected encoding of the input; this computation is \emph{independent} of 
the circuit being verified. For reasonably large inputs (see the row for \sjs\ in Table \ref{tabmuggles}), our GPU-based client implementation performed this computation over 100$\times$ faster than the
sequential implementation of \cite{itcs}.  For smaller inputs the speedup was unsurprisingly smaller due to increased overhead relative to total computation time. Still, we obtained a 15$\times$ speedup even for an input of length 65,536 ($256 \times 256$ matrix multiplication).

\medskip \noindent \textbf{Client-side speedup relative to unverifiable sequential algorithms.}
Our matrix-multiplication results clearly demonstrate that for problems requiring super-linear time to solve, even the sequential implementation
of \cite{itcs} will save the client time compared to doing the computation locally. Indeed, the runtime of the client is dominated by the cost of evaluating a single symbol in an error-corrected encoding of the input, and this cost grows linearly with the input size. Even for relatively small matrices of size $256 \times 256$, the client in the implementation of \cite{itcs} saved time. For matrices with tens of millions of entries, our results demonstrate that the client will still take just a few seconds, while performing the matrix multiplication computation would require orders of magnitude more time. Our results demonstrate
that GPU computing can be used to reduce the verifier's computation time by another 100$\times$.

\begin{figure}[t]
\begin{center}
\includegraphics[width=0.7\linewidth]{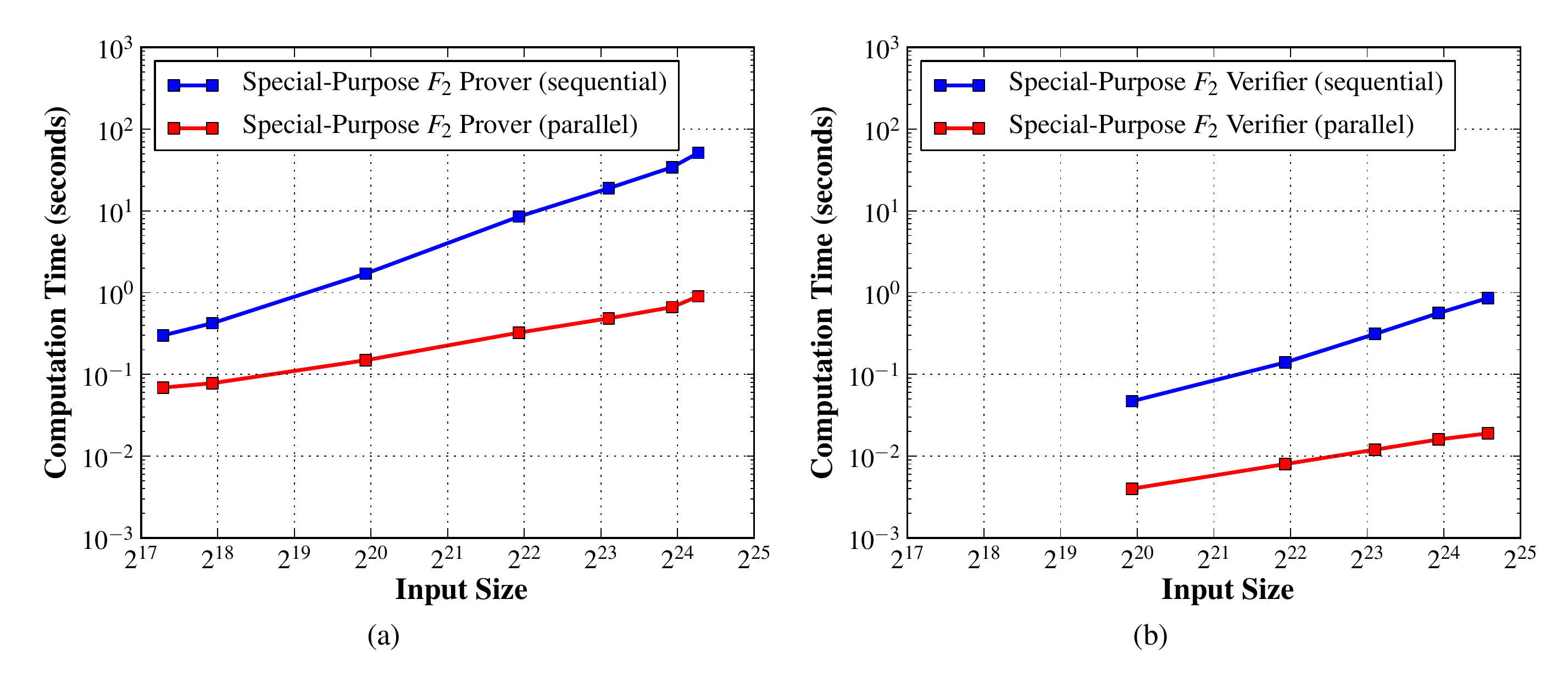}
\end{center}
\caption{Comparison of prover (a) and verifier (b) runtimes in the sequential and GPU-based implementations of the special-purpose $F_2$ protocol. Note that all plots are on a log-log scale.
Throughout, the verifier's space usage and the proof length are both set to $\sqrt{n}$.}
\label{fig:specialpurpose}
\end{figure}

\begin{table*}[t]
\centering
\begin{tabular}{ l l l l l l } 
\toprule
$\V$ space & Proof length & GPU $\P$  & Sequential $\P$  & GPU $\V$ & Sequential $\V$ \\
(KB)       & (KB)         & Time (s)  & Time (s)         & Time (s) & Time (s)        \\
\midrule
39.1   & 78.1   & 2.901 & 43.773 & 0.019 & 0.858  \\
78.2   & 39.1  & 1.872 & 43.544 & 0.010 & 0.639 \\
156.5    & 19.5  & 1.154 & 37.254 & 0.010 & 0.577 \\
313.2    & 9.8  & 0.909 & 36.554 & 0.008 & 0.552 \\
1953.1   & 0.78 & 0.357 & 20.658 & 0.007 & 0.551 \\
\bottomrule
\end{tabular}
\caption{Prover and verifier runtimes for the special-purpose $F_2$ protocol. All results are for fixed universe size $n=25$ million, varying the tradeoff between proof length and 
the client's space usage. This universe size corresponds to 190.7 MB of data.}
\label{tab:specialpurpose}
%\vspace*{-3mm}
\end{table*}

\subsection{Special-purpose protocols.}
We implemented both the client and the server of the non-interactive $F_2$ protocol of \cite{icalp09, itcs} on the GPU. 
As described in Section 2.3, this protocol is the fundamental building block for a host of non-interactive protocols achieving optimal tradeoffs between the space usage of the client and the length of the proof.
Figure \ref{fig:specialpurpose} demonstrates the performance of our GPU-based implementation of this protocol. Our GPU implementation obtained a 20-50$\times$ server-side speedup relative to the sequential implementation of \cite{itcs}. This speedup was only possible
after transposing the data grid into column-major order so as to achieve perfect memory coalescing, as described in Section \ref{sec:accesspatterns}. 
%This optimization eluded us for some time, as the data is more naturally processed in row-major order, as in the implementation of \cite{itcs}.

The server-side speedups we observed depended on the desired tradeoff between proof length and space
usage. That is, the protocol partitions the universe $[n]$ into a $v \times h$ grid
where $h$ is roughly the proof length and $v$ is the verifier's space usage. The prover
processes each row of the grid independently (many rows in parallel).
When $v$ is large, each row requires a substantial amount of processing. In this case, the overhead of parallelization is effectively amortized over the total computation time. If $v$ is smaller, then the overhead is less effectively amortized
and we see less impressive speedups. 

We note that Figure \ref{fig:specialpurpose} depicts the prover runtime for both the sequential implementation of \cite{itcs} and our GPU-based implementation with the parameters $h=v=\sqrt{n}$. 
With these parameters, our GPU-based implementation achieved roughly a 20$\times$ speedup relative to the sequential program. Table \ref{tab:specialpurpose} 
shows the costs of the protocol for fixed universe size $n=25$ million as we vary the tradeoff between $h$ and $v$. The data in this table shows that our parallel implementation enjoys a 40-60$\times$ speedup 
relative to the sequential implementation 
when $v$ is substantially larger than $h$. This indicates that we would see similar speedups even when $h=v=\sqrt{n}$ if we scaled to larger input sizes $n$. Notice that
universe size $n=25$ million corresponds to over 190 MBs of data, while the verifier's space usage and the proof length are hundreds or thousands of times smaller in all our experiments.
An unverifiable sequential algorithm for computing the second frequency moment over this universe required $0.031$ seconds; thus, our parallel server implementation achieved a slowdown of
10-100$\times$ relative to an unverifiable algorithm.

In contrast, the verifier's computation was much easier to parallelize, as its memory access pattern is highly regular. Our GPU-based implementation obtained 40-70$\times$ 
speedups relative to the sequential verifier of \cite{itcs} across all input lengths $n$, including when we set $h=v=\sqrt{n}$.

\section{Conclusions}
This paper adds to a growing line of work focused on obtaining fully practical methods for verifiable computation. Our primary contribution
in this paper was in demonstrating the power of parallelization, and GPU computing in particular, to obtain robust speedups in some
of the most promising protocols in this area. We believe the additional costs of obtaining correctness guarantees demonstrated in this paper 
would already be considered modest in many correctness-critical applications. 
Moreover, it seems likely that future advances in interactive proof methodology will also
be amenable to parallelization. This is because the protocols we implement utilize a number of common primitives 
(such as the \emph{sum-check protocol} \cite{sum-check}) as subroutines, and these primitives are likely to appear 
in future protocols as well.

Several avenues for future work suggest themselves. 
First, the GKR protocol is rather inefficient for the prover when applied to computations which are non-arithmetic in nature,
as the circuit representation of such a computation is necessarily large. Developing 
improved protocols for such problems (even special-purpose ones) would be interesting. Prime candidates include many graph problems
like minimum spanning tree and perfect matching.
More generally, a top priority is to further
reduce the slowdown or the memory-intensity for the prover in general-purpose protocols. Both these goals could be accomplished by
developing an entirely new construction that avoids the circuit representation of the computation; it is also possible
that the the prover within the GKR construction can be further optimized without fundamentally altering the protocol.

\label{sec:conclusion}

\bibliographystyle{abbrv}
\bibliography{biblio}

\end{document}